\documentclass[prc,preprint]{revtex4}
\begin{document}

\title{K$^+$N scattering data and exotic Z$^+$ resonances} 

\author{R.A. Arndt}
\author{I.I. Strakovsky}
\author{R.L. Workman}
\email{rworkman@gwu.edu}
\affiliation{Department of Physics, Center for Nuclear Studies, \\
         The George Washington University, Washington, D.C. 20052}
       
\begin{abstract}
Given the growing evidence for an exotic S=+1 resonance, seen in
kaon, photon and neutrino induced reactions, we reexamine the
existing K$^+$p and K$^+$d database in order to understand how
such a state could have been missed in previous studies. The
lack of structure in this database implies a width of an MeV
or less, assuming a state exists near 1540 MeV.
\end{abstract}

\maketitle

\section{A BRIEF HISTORICAL PERSPECTIVE}

Twenty years ago, many groups were involved in the study of
unconventional states. The 1984 edition of the PDG review contains
sections discussing Z$_0$ and Z$_1$ (I = 0, 1) K$^+$N resonances,
as well as dibaryons (B=2). The reported states had widths of 100 MeV or
more and came to be understood as `pseudo-resonances', structures
arising from the coupling to inelastic channels (N$\Delta$ in the
case of dibaryons, K$^*$N and K$\Delta$ in the case of Z-resonances).

Search the 2002 edition of the Review and you will only
see mention of exotic baryonic states in an obscure Non-qqq 
section of the Note on N and $\Delta$ resonances. In the area of
dibaryons, most recent studies have focused on narrow states, as
these cannot be so easily discounted as the broad states. The 
recently discovered Z resonance (now almost universally 
called the $\Theta (1540)$) has generated great interest
as it too is narrow, and the first determination of its 
mass and width (limit) agreed with a prediction of Diakonov,
Petrov, and Polyakov~\cite{dpp}.

\section{EVIDENCE FOR AN S=+1 STATE}

Two types of experiments now exist. Those (recent) measurements
claiming  structures
in their K$^+$n or K$^0$p mass distributions, and the (much
older) experiments, mainly K$^+$p and K$^+$d, which existed
prior to these measurements. Both may be valuable in understanding
the $\Theta$.

\subsection{Recent experiments}

Mass determinations for the $\Theta$ have been very
consistent, falling in the range 1540$\pm$10~MeV.
Width determinations so far have given just upper limits,
based on the experimental resolution. For the photon and
neutrino induced results, the width limit~\cite{exp} has been roughly
$\Gamma$ $<$ 20 MeV, while for the ITEP K$^+$Xe$\to$K$^0$pXe'
experiment, a limit of $\Gamma$ $<$ 9~MeV was given~\cite{itep}.

\subsection{Hints from the K$^+$p and K$^+$d database}

The K$^+$ scattering database shows remarkably little
evidence for a structure near 1540~MeV (corresponding to
a lab momentum of 440~MeV/c). An examination of the 
K$^+$d total cross sections led Nussinov to place a 
limit~\cite{nus} of $\Gamma$ $<$ 6~MeV on the $\Theta$ width. 
By re-analyzing the database, using the VPI KN PWA
code, a tighter limit of 1-2~MeV has been claimed~\cite{gw}.
Haidenbauer and Krein have added a narrow I=0
$J^P$ = ${1\over 2}^+$ $\Theta (1540)$ to their KN
meson-exchange model and conclude~\cite{hk} that either the width must
be considerably less than 5~MeV, or the resonance must
lie much closer to threshold. All of these estimates
are based on a general lack of structure at energies
corresponding to the $\Theta$.

These remarks also apply to the $\Theta ^{++}$ which
has been predicted to exist in order to have an
isotensor $\Theta$~\cite{cpr} and an isospin-violating
decay to KN, thus explaining the narrow width. 
No evidence is seen in the I=1 total cross sections
in the neighborhood of the $\Theta ^+$ energy.

\subsection{The Nussinov estimate reexamined}

Somewhat tighter limits on the width can be obtained
if the estimate of Nussinov is carefully
examined.  The limit $\Gamma$ $<$ 6~MeV is
based on there being 2-4~mb fluctuations in
the K$^+$d total cross sections for energies
corresponding to the $\Theta  (1540)$ mass.
The measurements of Carroll and Bowen cover 
this region, and both have data near 440~MeV/c.
Although there are normalization issues, the
deviations from a linear behavior are 
certainly less than 1~mb in both cases.
This alone reduces the Nussinov estimate
to $\Gamma$ $<$ 1.5~MeV.  [Deviations of
2-4~mb {\it are} visible near 600~MeV/c].

Further corrections to the Nussinov estimate have been
made by Cahn and Trilling~\cite{ct}. These include
the use of proper kinematic relations and a more
realistic treatment of the deuteron. These
modifications, together with an even more
conservative limit on fluctuations (1.5~mb)
lead to a width limit of less than an MeV.

\subsection{The Haidenbauer-Krein estimate}

A simple consideration of Clebsch-Gordon
coefficients shows that the effect of a
narrow I=0 resonance should be twice as
large in the I=0 total cross section, 
as compared to the K$^+$d total cross section.
This is evident in the Haidenbauer-Krein plot
of their model versus experimental data.

To be more realistic, however, the comparison
between models and I=0 data, extracted from
K$^+$d scattering measurements, should be modified.
The extraction process implicitly assumes
there are no sharp structures in the underlying
KN interactions. To compensate, it is therefore
more reasonable to apply the broadening
due to Fermi motion to the model result.
This still allows a strong limit.

\section{KN PARTIAL-WAVE ANALYSIS}

Our 1992 KN PWA analyzed both K$^+$p and K$^+$d
data~\cite{vpi}, finding broad counterclockwise motion in
the $P_{01}$, $D_{03}$, $P_{11}$, and $D_{15}$
Argand diagrams. No narrow structures were
reported. In order to search for missing
states, resonances with varying masses and
widths were inserted into the S-, P-, and
D-waves. These  waves were then refitted
to data, in order to see whether an improved
description was possible. 

For widths of 10~MeV or more, values in line
with the resolution-limited estimates, the
fit $\chi^2$ in some cases doubled, even
though the effect was localized and the fit
extended to 1.1~GeV or 2.65~GeV, depending
on the isospin. The fit remained worse 
until widths were reduced below the few MeV 
level, at which point a narrow structure
could fall into data gaps, having little
influence. 

We also examined the $\chi^2$ contributions
due to experiments closest to the resonance
position, as these values become more suspect
for very narrow widths. Discounting these
points, we continued to see no improvement
in the fit to the remaining database.

One should note that this test would not
be possible for a missing state above 
the inelastic threshold. In that case,
as in the $\pi$N
elastic scattering analysis, a missing
state could be understood in terms of a
small branching fraction to the elastic
channel.

\section{FUTURE WORK}

If the $\Theta (1540)$ exists and has a width of an MeV
or less, future measurements will have to carefully address the
problems associated with beam momentum uncertainty and
spread, as well as Fermi motion if deuteron targets are
used. The loss of meson beam facilities (including 
the associated manpower, expertise, and infrastructure) 
will make this task more difficult. 

It must be emphasized that we have only given upper limits
to the $\Theta$ width. It is not reasonable to calculate
coupling constants and make predictions based on a fixed
width of O(10~MeV).

\end{document}